\documentclass[journal,twocolumn,romanappendices]{IEEEtran}

\usepackage[T1]{fontenc}
\usepackage{url}
\usepackage{braket}
\usepackage{ifthen}
\usepackage{cite}
\usepackage[cmex10]{amsmath} 
\usepackage{longtable}
\usepackage{makecell}
\usepackage{algorithm,algorithmicx}
\usepackage[noend]{algpseudocode}
\usepackage{color}
\usepackage[normalem]{ulem}
                             
\usepackage{amssymb,amsthm}
\usepackage{mathtools,bbm}  
\usepackage{bm}
\usepackage{xifthen,xparse}
\usepackage{hyperref}  
\usepackage{float}
\usepackage{hhline}
\usepackage{caption}
\usepackage{subcaption}
\usepackage{rotating}

\usepackage{stmaryrd}

\usepackage{booktabs,tabularx}
\newcolumntype{C}{>{\centering\arraybackslash}X}

\usepackage[table]{xcolor}

\definecolor{dkgreen}{rgb}{0,0.6,0}
\definecolor{gray}{rgb}{0.5,0.5,0.5}
\definecolor{mauve}{rgb}{0.58,0,0.82}

\usepackage{pgfplots}
  \pgfplotsset{compat=newest}
  \usetikzlibrary{plotmarks}
  \usetikzlibrary{arrows.meta}
  \usepgfplotslibrary{patchplots}
  \usepackage{grffile}
  \usepackage{amsmath}

\usepackage{tikz,pgfplots,tkz-graph,color}
\pgfplotsset{compat=newest}
\pgfplotsset{plot coordinates/math parser=false}
\usetikzlibrary{decorations.markings,calc,positioning,matrix}
\usepackage{quantikz}

\usepackage{textcomp}

\allowdisplaybreaks

\newtheorem{example}{Example}

\newenvironment{example*}
  {\addtocounter{example}{-1}\example}
  {\endexample}

\newcommand{\llbr}{[\![}
\newcommand{\rrbr}{]\!]}

\newif\ifnotes
\notestrue

\begin{document}

\title{Tutorial on Quantum Error Correction for 2024 Quantum Information Knowledge (QuIK) Workshop}

 \author{%
   \IEEEauthorblockN{Priya J. Nadkarni, Narayanan Rengaswamy, and Bane Vasi{\'c}}%
    \thanks{P.~J. Nadkarni, N. Rengaswamy, and B. Vasi{\'c} are the Program Chairs for the First Quantum Information Knowledge (QuIK) Workshop held during the 2024 IEEE International Symposium on Information Theory at Athens, Greece.
    Email: narayananr@arizona.edu}
  }

{\maketitle}

\begin{abstract}
We provide a brief review of the fundamentals of quantum computation and quantum error correction for the participants of the first Quantum Information Knowledge (QuIK) workshop at the 2024 IEEE International Symposium on Information Theory (ISIT 2024).
While this is not a comprehensive review, we provide many references for the reader to delve deeper into the concepts and research directions.
\end{abstract}


\section{Introduction}
\label{sec:intro}

Quantum technologies exploit the laws of quantum mechanics, the most precise physical description of the world, to enable fundamentally new information processing capabilities.
The primary quantum technologies are quantum computers, quantum communications and networks, and quantum sensors.
While these technologies are all developed from the same concepts, their goals and tasks vary significantly.
For this workshop, we will primarily focus on quantum computing, where the goal is to store and process information in quantum-mechanically behaving carriers such as atoms, ions, superconducting circuits, and photons.
When isolated from their environments, these carriers behave ideally and can keep the information intact indefinitely.
However, in reality, they continuously interact with the environment and cause the stored information to decohere.
Similarly, the external manipulation of these carriers to compute on the information is also far from ideal, suffering from lack of precision, background noise etc.
Therefore, it is essential to protect the stored information from decoherence as well as ensure that its processing remains tolerant to faults in the apparatus.
The most systematic approach to such fault tolerant information processing in quantum systems is through the use of quantum error correcting codes.

In this document, we provide a brief overview of the fundamentals of quantum error correction and fault tolerance. 
We assume that the reader is familiar with classical error correction or channel coding but perhaps not with quantum information.
The goal is to provide sufficient background for the QuIK'24 workshop attendees to follow the invited talks, posters and discussions.
While this is not a comprehensive review of the field, we will provide ample references for the readers to expand on the fundamentals discussed here. 
For a historical review of quantum computing and quantum error correction, we recommend the readers to refer to~\cite{NielsenChuang, lidar2013quantum, QC_HistoricalReview_Shor, QC_HistoricalReview_Preskill}.
  
\section{Basics of Quantum Computation}

\subsection{Postulates of quantum mechanics}

The theory of quantum mechanics involves a mathematical formulation describing the behaviour of physical systems at submicroscopic scales with a set of postulates that associate experimental observations to the mathematical formulation. 
The four postulates of quantum mechanics are~\cite{NielsenChuang}:
\begin{enumerate}
 
\item \underline{State of a quantum mechanical system}: A normalized \textit{state vector}, a unit vector in the \textit{state space}, completely describes an isolated physical system. 
The state space is mathematically described by a Hilbert space, a complete complex vector space with an inner product.
The fundamental unit of quantum information is an $m$-dimensional quantum state called a \textit{quantum digit} (\textit{qudit}). 
When $m = 2$, the two-dimensional unit of quantum information in a two-level quantum system is termed the \textit{quantum bit} (\textit{qubit}), whose state is represented by the ``ket''
\begin{align}
\ket{\psi} = a\ket{0} + b\ket{1} = a \begin{bmatrix} 1 \\ 0 \end{bmatrix} + b \begin{bmatrix} 0 \\ 1 \end{bmatrix}, \label{eqn:State_Rep}
\end{align} 
where $a,b \in \mathbb{C}$ and $|a|^2+|b|^2=1$.
The normalization constraint is referred to as \emph{Born's rule}~\cite{wilde2013quantum}.
The states $\ket{0}$ and $\ket{1}$ form the \textit{computational basis} of the state space. 
The state $\ket{\psi}$ is said to be in a \emph{superposition} of $\ket{0}$ and $\ket{1}$. 
The Hermitian transpose of a ket is the ``bra'': 
\begin{align}
\bra{\psi} \coloneqq \ket{\psi}^{\dagger} &= a^* \bra{0} + b^* \bra{1} \nonumber \\
  &= a^* \begin{bmatrix} 1 & 0 \end{bmatrix} + b^* \begin{bmatrix} 0 & 1 \end{bmatrix}.
\end{align}
The (complex) inner product between two quantum states $\ket{\psi}$ and $\ket{\phi}$ is denoted by $\braket{\psi \vert \phi}$.
The \emph{bra-ket} notation is also termed as the \emph{Dirac} notation, named after Paul Dirac who introduced it%
\footnote{His intention was likely to make the inner product $\braket{\psi \vert \phi}$ look similar to the common bracket notation $(\ket{\psi},\ket{\phi})$, but specific to quantum mechanics.}.
Similar to qubits, the state of an $m$-dimensional qudit is represented by $\ket{\psi} = \underset{j=0}{\overset{m-1}{\sum}} a_j \ket{j}$, where $a_0,a_1,\dots,a_{m-1} \in \mathbb{C}$ and $\sum_{j=0}^{m-1} |a_j|^2 = 1$.
The states $\ket{0}$, $\ket{1}$, $\dots$, $\ket{m-1}$ form the computational basis of the qudit state space.

\item \underline{Evolution of a quantum mechanical system}: The evolution of a closed (or isolated) quantum system is completely described by a unitary operator. 
Recall that a complex square matrix $U \in \mathbb{C}^{2 \times 2}$ is unitary if and only if its inverse is the same as its Hermitian transpose, i.e., $U^{-1} = U^{\dagger}$.
The states $\ket{\psi_1}$ and $\ket{\psi_2}$ of a quantum system at times $t_1$ and $t_2$ are related by a unitary operator $U$ that depends only on the time instances $t_1$ and $t_2$, i.e., $\ket{\psi_2} = U\ket{\psi_1}$. 
The most basic unitary operators are the single-qubit \emph{Pauli} operators
\begin{align}
\mathrm{I}_2 = 
\begin{bmatrix}
 1 & 0 \\
 0 & 1
\end{bmatrix} \ &, \  
\mathrm{X} = 
\begin{bmatrix}
 0 & 1 \\
 1 & 0
\end{bmatrix},  \nonumber \\
\mathrm{Y} = 
\begin{bmatrix}
 0 & -\mathrm{i}\\ 
 \mathrm{i} & 0
\end{bmatrix} \ &, \   
\mathrm{Z} = 
\begin{bmatrix}
 1 & 0 \\
 0 & -1
\end{bmatrix}.
\end{align} 
Note that these are also Hermitian and have unit determinant.
The non-identity Pauli operators have order $2$, zero trace, and eigenvalues $\pm 1$.
These generate the single-qubit Pauli group
\begin{align}
\mathcal{P} & \coloneqq \langle \, \mathrm{i} \mathrm{I}_2, \mathrm{X}, \mathrm{Y}, \mathrm{Z} \, \rangle \\
  &= \left\{ \pm \mathrm{I}_2, \pm\mathrm{i}\mathrm{I}_2,\pm \mathrm{X}, \pm\mathrm{i}\mathrm{X},\pm \mathrm{Y}, \pm\mathrm{i}\mathrm{Y},\pm \mathrm{Z}, \pm\mathrm{i}\mathrm{Z}\right\}.
\end{align}

The Pauli matrices are also represented as $\sigma_0 \equiv \mathrm{I_2}$, $\sigma_1 \equiv \mathrm{X}$, $\sigma_2 \equiv \mathrm{Y}$, and $\sigma_3 \equiv \mathrm{Z}$. 
Two non-identity Pauli matrices anticommute with each other, e.g., $\mathrm{X} \mathrm{Z} = - \mathrm{Z} \mathrm{X}$, and are related as $\mathrm{XY}=\mathrm{iZ},\mathrm{YZ}=\mathrm{iX},\text{ and }\mathrm{ZX}=\mathrm{iY}$. 
The Pauli matrices are orthogonal with respect to the Hilbert-Schmidt (or trace) inner product $\langle A,B \rangle_{\rm HS} \coloneqq \mathrm{Tr}(A^{\dagger}B)$. 
They form an orthonormal basis via normalization: 
\begin{align}
\left\{ \frac{1}{\sqrt{2}} \sigma_i \ ; \ i=0,1,2,3 \right\}.
\end{align}
The Hermitian Pauli matrices are represented by a two-bit vector $[a,b]$ based on the representation of the operator as $\mathrm{i}^{ab} \mathrm{X}^a\mathrm{Z}^b$, i.e., $\mathrm{I} \equiv [0,0], \mathrm{X} \equiv [1,0], \mathrm{Y} \equiv [1,1], \mathrm{Z} \equiv [0,1]$. 
By extension to $\mathcal{P}$, this defines a homomorphism $\gamma \colon \mathcal{P} \rightarrow \mathbb{F}_2^2$ whose kernel is $\langle \, \mathrm{i} \mathrm{I_2} \, \rangle$.

\quad Similarly, the evolution over qudits are described by unitary operators in $\mathbb{C}^{m \times m}$, which are usually represented in terms of the generalized Pauli basis~\cite{WeylHeisenberg}. 
For qudits of prime-power dimension $m=p^l$, the unitary operators are represented in terms of the clock operators $\mathrm{Z}^{(p^l)}(\gamma) = \sum_{\theta \in \mathbb{F}_{p^l}}\omega^{\mathrm{Tr}_{p^l/p}(\gamma\theta)}\ket{\theta}\bra{\theta}$ and shift operators $ \mathrm{X}^{(p^l)}(\beta) = \sum_{\theta \in \mathbb{F}_{p^l}}\ket{\beta + \theta}\bra{\theta}$, where $p$ is prime, $\omega\!=\!\mathrm{e}^{\frac{\mathrm{i}2\pi}{p}}$, $\beta \!\in\! \mathbb{F}_{m}$, and the field trace $\mathrm{Tr}_{p^l/p}(\beta)\!=\!\sum^{l-1}_{i=0}\beta^{p^l}$~\cite{Ketkar06 , NBEASC2021}. 

\item \underline{Measurement on a quantum mechanical system}: A set of operators $\{ M_m \}_m$ satisfying $\sum_m M_m^{\dagger} M_m = I$, called the \textit{measurement operators}, describe a quantum measurement, where the index $m$ denotes the possible measurement outcomes. 
The measurement outcome and the post-measurement state are probabilistic in nature: if the state of the system being measured is $\ket{\psi}$, then the outcome $m$ is obtained with probability 
\begin{align}
p(m) = \bra{\psi} M_m^{\dagger}M_m \ket{\psi}. 
\end{align}
The \emph{completeness condition} $\sum_m M_m^{\dagger} M_m = I$ ensures that the probabilities sum to $1$ for any initial state $\ket{\psi}$.
The post-measurement state of the system is given by
\begin{align}
\ket{\psi_m} = \frac{M_m \ket{\psi}}{\sqrt{p(m)}} = \frac{M_m \ket{\psi}}{\sqrt{ \bra{\psi} M_m^{\dagger}M_m \ket{\psi} }}.
\end{align}
Thus, measurement destroys superposition unless the state is an eigenstate of a measurement operator.

\quad The most common measurement is a \emph{projective} measurement, where $M_m = P_m$ are projection operators satisfying $P_m P_{m'} = 
\begin{cases}
P_m & \text{if}\ m = m', \\
0 & \text{if}\ m \neq m'.
\end{cases}$
It is common to describe a projective measurement as the measurement of an \emph{observable}, i.e., a Hermitian operator.
The outcomes are the eigenvalues and the measurement operators are given by the projectors onto the different eigenspaces, obtained by diagonalizing the observable.
As an example, consider the measurement of $\mathrm{Z}$ on $\ket{\psi} = a \ket{0} + b \ket{1}$.
It is easy to verify that the projectors are $P_{+1} = \ket{0}\bra{0}, P_{-1} = \ket{1}\bra{1}$.
The probabilities (resp. post-measurement states) of outcomes $+1$ and $-1$ are $|a|^2$ and $|b|^2$ (resp. $\ket{0}$ and $\ket{1}$), respectively.

\quad It is important to note that no quantum measurement can distinguish a state $\ket{\psi}$ from its scalar multiple $e^{\mathrm{i} \theta} \ket{\psi}$.
Hence, \emph{global} phase never matters.

\item \underline{Composite quantum mechanical systems}: The state space of a composite physical system is the tensor product of the state spaces of the component physical systems. 
For example, the state space of an $n$-qubit system is $\mathbb{C}^{2^n} = \mathbb{C} \otimes \mathbb{C} \otimes \cdots \otimes \mathbb{C}$.
For a composite physical system with the $i^{\rm th}$ system is prepared in state $\ket{\psi_i}$, where $i \in \{1,\ldots,n\}$, the state of the complete system is $\ket{\psi_1} \otimes \ket{\psi_2} \otimes \cdots \otimes \ket{\psi_n} \in \mathbb{C}^{2^n}$.
The evolution of a closed quantum system with $n$ qubits is described by a unitary operator $U \in \mathbb{C}^{2^n \times 2^n}$.
The Pauli group $\mathcal{P}_n$ on an $n$-qubit system is defined as the $n$-fold tensor product of the single-qubit Pauli group $\mathcal{P}$.
The homomorphism $\gamma$ is extended to map $\mathcal{P}_n$ to $\mathbb{F}_2^{2n}$: 
\begin{align} 
& E(\bm{a}, \bm{b}) \coloneqq \mathrm{i}^{a_1 b_1} \mathrm{X}^{a_1} \mathrm{Z}^{b_1} \otimes \cdots \otimes \mathrm{i}^{a_n b_n} \mathrm{X}^{a_n} \mathrm{Z}^{b_n} \nonumber \\ 
& \mapsto [a_1, a_2, \ldots, a_n, b_1, b_2, \ldots, b_n].
\end{align}
Since any two non-identity Pauli matrices anti-commute, we can define the \emph{symplectic inner product} between their binary representations to capture commutativity:
\begin{align}
\mathrm{symp}([\bm{a},\bm{b}],[\bm{c},\bm{d}]) \coloneqq \bm{c}\bm{b}^T + \bm{a}\bm{d}^T (\bmod\ 2).
\end{align}
The corresponding Pauli operators $E(\bm{a}, \bm{b})$ and $E(\bm{c}, \bm{d})$ commute if and only if the above symplectic inner product is $0$~\cite{rengaswamy2020classical}.
The $2^{2n}$ Hermitian $n$-qubit Pauli matrices 
\begin{align}
\left\{ \frac{1}{\sqrt{2^n}} E(\bm{a},\bm{b}) \ ; \ \bm{a}, \bm{b} \in \mathbb{F}_2^n \right\}
\end{align}
form an orthonormal basis under the Hilbert-Schmidt inner product. Similarly, an $n$-qudit system can be defined as an $n$-fold tensor product of the single-qudit generalized Pauli group~\cite{NBEASC2021}.
\end{enumerate}

\subsection{Mixed states and entanglement}

\textit{Entanglement} is a physical phenomenon unique to quantum mechanics due to which the definite state of a component system of a composite physical system cannot be described independently of the other component systems, irrespective of the distance between them. 
Entanglement and superposition enable quantum systems to perform better compared to their classical counterparts. 
For every quantum system in state $\ket{\psi}$ with sub-systems A and B, there exist a \textit{Schmidt decomposition} in terms of the orthonormal basis states $\{ \ket{j_A} \}_j$ for the sub-system A and orthonormal basis states $\{ \ket{j_B} \}_j$ for the sub-system B, respectively, such that
$\ket{\psi} = \underset{j=1}{\overset{r}{\sum}}\lambda_j\ket{j_A}\ket{j_B}$,
where $\lambda_j$s are non-negative numbers called the \emph{Schmidt coefficients} such that $\underset{j=1}{\overset{r}{\sum}}\lambda_j^2=1$ and $r$ is called the \emph{Schmidt rank}~\cite{NielsenChuang , wilde2013quantum}.
The state $\ket{\psi}$ is entangled if and only if $r > 1$.

In an entangled system, the sub-system is viewed to be in an ensemble of states $\{(p_i,\ket{\psi_i})\}_i$, meaning that the sub-system is in the state $\ket{\psi_i}$ with probability $p_i$. 
When the ensemble contains only one element, the sub-system is said to be in a \textit{pure state}; else, it is in a \textit{mixed state}. 
The state of the sub-system can alternatively be represented by a \textit{density matrix} $\rho = \underset{i}{p_i\ket{\psi_i}\bra{\psi_i}}$, which is a positive operator whose trace is $1$. 
For a pure state, $\mathrm{Tr}(\rho^2) = 1$, while for a mixed state, $\mathrm{Tr}(\rho^2) < 1$. 
An example of an entangled state is 
\begin{align}
\frac{\ket{00} + \ket{11}}{\sqrt{2}} \equiv \frac{\ket{0} \otimes \ket{0} + \ket{1} \otimes \ket{1}}{\sqrt{2}}, 
\end{align}
represented by the ensemble $\left\{ \left( 1, \frac{\ket{00} + \ket{11}}{\sqrt{2}} \right) \right\}$. 
However, each of its single-qubit sub-systems are described by the ensemble $\{(\frac{1}{2},\ket{0}), (\frac{1}{2},\ket{1})\}$ and, hence, the density matrix $\frac{1}{2}\mathrm{I}_2$. 
This is one of the four famous \emph{Bell states} or \emph{EPR pairs} (EPR stands for Einstein-Podolsky-Rosen): 
\begin{align}
\ket{\Phi^{\pm}} = \frac{\ket{00} \pm \ket{11}}{\sqrt{2}} \ , \ \ket{\Psi^{\pm}} = \frac{\ket{01} \pm \ket{10}}{\sqrt{2}}.     
\end{align}
Together, these form an entangled basis for $\mathbb{C}^4$, the state space of two qubits.
They play a critical role in quantum information.

We note that two different ensembles could correspond to the same density matrix, where the states of one ensemble are linear combinations of states of the other and the coefficients of the linear combinations form a unitary matrix~\cite{NielsenChuang}.
According to the four postulates of quantum mechanics, the unitary evolution of a mixed state $\rho$ under a unitary operator $U$ is described by $U\rho U^{\dagger}$, where $U^{\dagger}$ is the conjugate transpose of $U$. 
Performing a measurement on $\rho$ described by measurement operators $\{M_m\}_m$ collapses $\rho$ to the state 
\begin{align}
\rho_m = \frac{M_m\rho M_m^{\dagger}}{\mathrm{Tr}(M_m^{\dagger}M_m\rho)} 
\end{align}
with probability $p(m) = \mathrm{Tr}(M_m^{\dagger}M_m\rho)$.
Sometimes, it suffices to know only the statistics of the measurement and not the post-measurement states.
For such cases, we can define a \emph{POVM (positive operator-valued measure)} using positive operators $\{ E_m \}_m$ such that $\sum_m E_m = I$, where $E_m$ plays the role of $M_m^{\dagger} M_m$ and $p(m) = \text{Tr}(E_m \rho)$.

\subsection{Quantum gates and measurements}

A quantum gate is the same as a unitary operator.
The basic gates are the Pauli gates and their tensor products on $n$ qubits.
An important set of gates is the \emph{Clifford group} $\mathcal{C}_n$~\cite{bennett1996mixed , calderbank1997quantum , Gottesman97 , Calderbank_GF4_97}, which is defined as the \emph{normalizer} of the Pauli group $\mathcal{P}_n$, i.e.,
\begin{align}
\mathcal{C}_n \coloneqq \{ U \in \mathbb{U}^{2^n} \colon U P U^{\dagger} \in \mathcal{P}_n \ \forall \ P \in \mathcal{P}_n \},
\end{align}
where $\mathbb{U}^{2^n}$ denotes the group of unitary matrices of size $2^n$.
In other words, Clifford gates conjugate Pauli gates to Pauli gates.
The Clifford group is generated by three gates: Hadamard ($\mathrm{H}$), Phase ($\mathrm{S}$), and controlled-NOT ($\mathrm{CNOT}$).
Their matrix representations are provided below:
\begin{align}
\mathrm{H} &\coloneqq \frac{1}{\sqrt{2}}
\begin{bmatrix}
1 & 1 \\
1 & -1
\end{bmatrix} \ , \ 
\mathrm{S} \coloneqq 
\begin{bmatrix}
1 & 0 \\
0 & \mathrm{i}
\end{bmatrix} = \sqrt{\mathrm{Z}} \ , \nonumber \\
\mathrm{CNOT} &= 
\begin{bmatrix}
1 & 0 & 0 & 0 \\
0 & 1 & 0 & 0 \\
0 & 0 & 0 & 1 \\
0 & 0 & 1 & 0
\end{bmatrix} =
\begin{bmatrix}
\mathrm{I}_2 & 0 \\
0 & \mathrm{X}
\end{bmatrix} \nonumber \\
  &= \ket{0}\bra{0} \otimes I_2 + \ket{1}\bra{1} \otimes \mathrm{X} \equiv \mathrm{CX}.
\end{align}
Here, the notation $\mathrm{CX}$ refers to the fact that $\mathrm{CNOT}$ is a ``controlled-$\mathrm{X}$'' gate: if the first qubit (\emph{control}) is in state $\ket{0}$, then it does nothing to the second qubit (\emph{target}), but if the control qubit is in state $\ket{1}$, then it applies $\mathrm{X}$ to the target qubit.
Since $\mathrm{X}$ is the ``bit flip'' gate, i.e., $\mathrm{X} \ket{0} = \ket{1}$ and $\mathrm{X} \ket{1} = \ket{0}$, the effect of CNOT is the same as the reversible XOR generalized to quantum states via linearity.
Note that $\mathrm{Z}$ is commonly called the ``phase flip'' gate, since $\mathrm{Z} \ket{0} = \ket{0}$ and $\mathrm{Z} \ket{1} = - \ket{1}$, and $\mathrm{Y}$ is called the ``bit-phase flip'' gate.

The Clifford group can be extended to a \emph{universal} gate set by including any non-Clifford gate.
Here, universality means that any unitary operator can be decomposed into a sequence of gates from this finite set with arbitrarily small approximation error in the spectral norm~\cite{NielsenChuang , boykin2000new}.
The most common non-Clifford gate included in the universal set is 
\begin{align}
\mathrm{T} \coloneqq 
\begin{bmatrix}
1 & 0 \\
0 & e^{\mathrm{i} \frac{\pi}{4}}
\end{bmatrix} = \sqrt{\mathrm{S}} = \mathrm{Z}^{\frac{1}{4}},
\end{align}
called the ``$\mathrm{T}$ gate''.
Hence, a common universal gate set for quantum computing is $\{ \mathrm{H}, \mathrm{T}, \mathrm{CNOT} \}$.
The circuit notations for single-qubit gates $U$ and $\mathrm{CNOT}_{1 \rightarrow 2}$ (i.e., first qubit as control and second qubit as target) are shown below:
\begin{center}
\begin{quantikz}
\lstick{$\ket{\psi}$} & \gate{U} & \rstick{$U \ket{\psi}$}
\end{quantikz} 
\quad
\begin{quantikz}
\lstick[wires=2]{$\ket{\psi}$} & \ctrl{1} & \rstick{control} \\
 & \targ{} & \rstick{target}
\end{quantikz}
\end{center}

The only other ingredient in quantum circuits is the quantum measurement.
It can be shown that general quantum measurements can be realized through additional ancillary qubits, joint unitary evolution, and projective measurements~\cite{NielsenChuang}.
Hence, it suffices to only consider projective measurements in quantum circuits.
The standard measurement is the measurement of Pauli $\mathrm{Z}$, often called as the $\mathrm{Z}$-basis measurement.
Other measurements can be realized through suitable unitary operations before $\mathrm{Z}$-measurement, e.g., $\mathrm{X}$-measurement is equivalent to applying $\mathrm{H}$ followed by $\mathrm{Z}$-measurement since $\mathrm{H} \mathrm{Z} \mathrm{H}^{\dagger} = \mathrm{X}$.
The circuit representation for such measurements are:
\begin{center}
\begin{quantikz}
\lstick{$\ket{\psi}$} & \meter{\mathrm{X}} & \setwiretype{c} \rstick{$\pm 1$}
\end{quantikz}
$=$
\begin{quantikz}
\lstick{$\ket{\psi}$} & \gate{\mathrm{H}} & \meter{\mathrm{Z}} & \setwiretype{c} \rstick{$\pm 1$}
\end{quantikz}
\end{center}
The double wire represents classical information whereas a solid wire represents quantum information.

\section{Physical Realization of Qubits}

Most of the quantum computing systems currently use qubits. 
The physical implementation of these qubits can be based on various technologies such as photonics, superconducting circuits, ion traps, quantum dots, neutral atoms, etc.~\cite{NielsenChuang,PreskillNotes}. 
At the moment, there is no particular technology considered the standard for implementation of quantum computers, unlike classical computers for which semiconductor technology is considered the standard. 

Photonic qubit encodings are usually based on using either a photon's degree of freedom, such as its polarization, or using continuous-variable codes, such as bosonic codes, based on states of light to encode a qubit~\cite{XanaduBlueprint , slussarenko2019photonic , romero2024photonic , killoran2019strawberry , kolarovszki2024piquasso}. 
Photonic quantum computers are easy to network, usually have minimal cryogenics requirement, are scalable, have flexbility in choice of quantum error correction code used, and use measurement-based quantum computing (MBQC)~\cite{raussendorf2001one , raussendorf2003measurement , raussendorf2006fault , briegel2009measurement} approaches. 
Their main challenge is the probabilistic photonic-based qubit generation/gates and to combat photon loss.
 
Superconducting qubit encodings use superconducting electronic circuits to encode qubits within artificial atoms~\cite{SuperconductingQCReview , rasmussen2021superconducting , kwon2021gate , bravyi2022future}. 
The basis states of a qubit are mapped to the energy levels that correspond to the integer number of pairs of electrons called Cooper pairs (for charge qubits), or to the integer number of magnetic flux quanta (for flux qubits), or different charge oscillation amplitudes across a Josephson junction (for phase qubits/qudits)~\cite{Superconducting_Qudit18}. 
Superconducting qubits have fast gate times and methods/processes used for implementing classical computers can be utilized. 
However, their architectures need to be designed with quantum error correction codes whose operations act on neighboring qubits as the qubits are usually laid out on a surface and have only limited nearest-neighbor coupling. 
Due to this restriction, scaling on superconducting quantum systems is a challenge.
 
Ion-trap qubit encodings use ions or charged particles confined and suspended in free space using electromagnetic fields~\cite{IonTrapReview , bernardini2023quantum , moses2023race}.
The basis states are the stable energy levels of these ions. 
Ion-trap qubits have long coherence times and high-fidelity quantum operations. 
The main challenge for ion-trap based quantum computing is scaling it to hundreds or thousands of qubits, which is required for quantum advantage. 
 
Neutral atom encodings use two different energy states of the atom to encode the qubit \cite{NeutralAtomReview , wurtz2023aquila , young2022architecture}. 
The atoms have long coherence time and are easier to trap and control as they are neutral in charge, enabling scalable quantum computing architectures. 
The gate operations are slower compared to  superconducting circuits and need more preparation time at the beginning of the computation.

\section{Quantum Noise Channels}

A general representation of a quantum channel is the so-called \emph{Kraus representation}~\cite{NielsenChuang , wilde2013quantum}:
\begin{align}
\mathcal{E}(\rho) = \sum_i E_i \rho E_i^{\dagger},
\end{align}
where $\{ E_i \}_i$ are called Kraus operators and satisfy the completeness condition $\sum_i E_i^{\dagger} E_i = I$.
The effect of noise on a quantum system can be captured in this representation using suitable Kraus operators that describe the noise.
The \emph{dephasing channel} either leaves the qubit unchanged or applies $\mathrm{Z}$:
\begin{align}
\mathcal{E}_{\rm dephasing}(\rho) = (1-\varepsilon) \ \rho + \varepsilon \ \mathrm{Z} \, \rho \, \mathrm{Z}.
\end{align}
The \emph{bit flip} channel can be described similarly as
\begin{align}
\mathcal{E}_{\rm flip}(\rho) = (1-\varepsilon) \ \rho + \varepsilon \ \mathrm{X} \, \rho \, \mathrm{X}.
\end{align}
The quantum equivalent of the classical binary symmetric channel is the \emph{depolarizing channel} which either leaves the qubit unchanged or applies one of the flip operators:
\begin{align}
\mathcal{E}_{\rm dep}(\rho) &= (1-\varepsilon) \ \rho + \frac{\varepsilon}{3} \ \left[ \mathrm{X} \, \rho \, \mathrm{X} + \mathrm{Y} \, \rho \, \mathrm{Y} + \mathrm{Z} \, \rho \, \mathrm{Z} \right] \\
  &= \left( 1 - \frac{4\varepsilon}{3} \right) \ \rho + \frac{4 \varepsilon}{3} \ \mathrm{I}_2.
\end{align}
The second equality can be shown by expanding $\rho$ in the Pauli basis~\cite[Chapter 3]{PreskillNotes}.
This is the most common noise channel considered in quantum computing as it represents the worst-case scenario where the quantum state can be replaced with the completely mixed state $\frac{1}{2} \mathrm{I}_2$ (which retains no information about $\rho$).
It can be simulated by a $4$-sided coin flip and applying either $\mathrm{I}_2, \mathrm{X}, \mathrm{Y}$, or $\mathrm{Z}$ according to the result.

\section{Quantum Error Correction}

Quantum error correction (QEC) involves incorporating redundancy in quantum information which enables the system to retrieve the quantum information in the presence of noise. 
The \emph{no cloning theorem}~\cite{NielsenChuang} forbids copying of arbitrary quantum states, so a na{\"i}ve quantum repetition code does not exist.
The QEC code is a subspace of the state space over which the quantum states are defined. 
A code is able to correct a set of errors $E = \{E_i\}$ if and only if $PE_i^{\dagger}E_jP = c_{ij} P$, where $P$ is the code space projector and $c_{ij} \in \mathbb{C}$ form a Hermitian matrix~\cite{NielsenChuang}. 
This is known as the \emph{Knill-Laflamme} condition for QEC~\cite{KnillLaflamme}. 
The \emph{stabilizer} framework~\cite{Gottesman97 , Calderbank_GF4_97} based on the Pauli group is commonly used to construct quantum codes. 
The \emph{Calderbank-Shor-Steane (CSS)} framework~\cite{CSS_CS , CSS_Steane}, a sub-framework of the stabilizer framework, constructs quantum codes from pairs of classical codes satisfying a dual-containing constraint. 
In this section, we review these two frameworks and discuss some of the latest codes of interest in the field of quantum computation.
The framework of \emph{subsystem codes}~\cite{kribs2005unified , poulin2005stabilizer , aly2006subsystem , bacon2006operator , breuckmann2011subsystem} generalizes the stabilizer framework and has proven very useful, but we will not discuss them here.

\subsection{Stabilizer codes}

Let $\mathcal{S}$ be an abelian subgroup of the Pauli group $\mathcal{P}_n$ that does not contain $-\mathrm{I}_{2^n}$. 
Let the minimal generators of $\mathcal{S}$ be $S_1$, $S_2$, $\dots$, $S_r$. 
The stabilizer code~\cite{Gottesman97 , Calderbank_GF4_97} $\mathcal{Q}_{\mathcal{S}}$ is the subspace of $\mathbb{C}^{2^n}$ defined as
\begin{align}
\mathcal{Q}_{\mathcal{S}} \coloneqq \left\{ \ket{\psi} \in \mathbb{C}^{2^n} \colon S_i \ket{\psi} = \ket{\psi} \ \forall \ i \in \{1,\dots,\rho\} \right\}.
\end{align}
The group $\mathcal{S}$ is called the \textit{stabilizer group} of $\mathcal{Q}_{\mathcal{S}}$ because it stabilizes the codeword $\ket{\psi}$, i.e., $\ket{\psi}$ is a simultaneous eigenstate of all elements of $\mathcal{S}$ with eigenvalue $+1$. 
The minimal generators $S_i$s are called the \textit{stabilizer generators}. 
Let $S_i = E(\bm{a}_i, \bm{b}_i)$,
where $\bm{a}_i, \bm{b}_i \in \mathbb{F}_2^n$.
The check matrix of $\mathcal{Q}_{\mathcal{S}}$ is 
$H_{\mathcal{S}}  = \big[ H_{\mathrm{X}} \big| H_{\mathrm{Z}} \big]$, where 
\begin{align}
H_{\mathrm{X}} = 
\begin{bmatrix}
\bm{a}_1 \\ \bm{a}_2 \\ \vdots \\ \bm{a}_r
\end{bmatrix} \ , \ 
H_{\mathrm{Z}} = 
\begin{bmatrix}
\bm{b}_1 \\ \bm{b}_2 \\ \vdots \\ \bm{b}_r
\end{bmatrix} \in \mathbb{F}_2^{r \times n}.
\end{align}
As the stabilizers commute, their symplectic inner product $\mathrm{symp}([\bm{a_i},\bm{b_i}],[\bm{a_j},\bm{b_j}]) = \bm{a_j}\bm{b_i}^{\mathrm{T}} + \bm{a_i}\bm{b_j}^{\mathrm{T}} = 0$ (mod $2$) for all $i,j\in \{1,2,\ldots,r\}$~\cite{rengaswamy2020classical}. 
Equivalently, we have the constraint 
\begin{align}
H_{\mathrm{X}} H_{\mathrm{Z}}^T + H_{\mathrm{Z}} H_{\mathrm{X}}^T = 0.
\end{align}

The dimension of the stabilizer code defined by $r$ minimal stabilizer generators over $n$ qubits is $2^{(n-r)}$~\cite{Gottesman97,Ashikhmin01,Ketkar06}. 
The code is said to encode $k = n-r$ \emph{logical qubits} of information into $n$ \emph{physical qubits}.
The normalizer $\mathcal{N}(\mathcal{S})$ of $\mathcal{S}$ in $\mathcal{P}_n$ is the set of operators in $\mathcal{P}_n$ that commute with the group $\mathcal{S}$, i.e., $\forall ~E \in \mathcal{N}(\mathcal{S})$ and $S \in \mathcal{S}$, we have $S E = E S$. 
The commonly considered errors on stabilizer codes are Pauli errors, e.g., the depolarizing channel.
Through appropriate syndrome measurement circuits that only depend on the stabilizers, one can detect if the error commutes or anti-commutes with each stabilizer~\cite{NielsenChuang}.
This provides an $r$-bit syndrome, where the bit is $0$ if the error commutes with that stabilizer and $1$ if it anti-commutes with that stabilizer.
Note that errors that are stabilizers leave the state unchanged and, hence, are trivial errors.
These are called \emph{degenerate} errors.
The minimum distance of the code, $d$, is the minimum weight of an undetectable error, i.e., a non-trivial error whose syndrome is trivial. 
Hence, $d$ is the minimum Pauli weight of an element in $\mathcal{N}(\mathcal{S}) \setminus \mathcal{S}$, where Pauli weight refers to the number of non-identity components in the $n$-qubit Pauli operator.
The size of $\mathcal{N(S)}$ is $2^{(2n-r)}$.
Overall, the stabilizer code has parameters $\llbr n,k,d \rrbr$.
 
The stabilizer codes can be viewed to be analogous to classical additive\footnote{All additive codes over a prime field are linear codes.} codes~\cite{Calderbank_GF4_97}. 

The check matrix is analogous to the parity check matrix of a classical code. 
As measurement of a quantum state collapses the superposition of the state, quantum error correction needs to be performed without any knowledge about the state. 
A syndrome is an $r$-bit binary vector obtained using the eigenvalues of the stabilizers for the erroneous quantum state, mapping $+1$ to bit $0$ and $-1$ to bit $1$. 
In other words, if $E \in \mathcal{P}_n$ is the error and $S \in \mathcal{S}$, then for any initial code state $\ket{\psi}$ we have the eigenvalue equation
\begin{align}
S (E \ket{\psi}) &= (SE) \ket{\psi} \\
 &= (\pm ES) \ket{\psi} \\ 
 &= (\pm E) (S \ket{\psi}) \\
 &= \pm E \ket{\psi}.
\end{align}
Hence, mathematically, the syndrome is obtained from the symplectic inner product of the error with the stabilizer generators. 
This is analogous to obtaining a syndrome in the classical case based on the parity checks. 
Based on the syndrome, a recovery operator is deduced and used to correct the error. 
Due to the existence of degenerate errors in the quantum setting, it suffices to find a recovery operator that is a product of the actual error and any stabilizer.
Thus, degeneracy is a uniquely quantum phenomenon which provides many ways to correct the same error.
If $E$ is the actual error and $\hat{E}$ is the error estimate from a decoder (that has the same syndrome as $E$), then there are two possible scenarios: $\hat{E} E \in \mathcal{S}$ (correct decoding) or $\hat{E} E \in \mathcal{N}(\mathcal{S}) \setminus \mathcal{S}$ (logical error).
A stabilizer code is said to be degenerate if there exists a stabilizer in $\mathcal{S}$ whose Pauli weight is strictly less than $d$.
For a degenerate code, correct decoding may be possible with $\hat{E} \neq E$ such that both $\hat{E}$ and $E$ have the same Pauli weight.

Qudit stabilizer framework is a generalization of the qubit stabilizer framework where the quantum code is also simultaneously stabilized by an abelian group. 
The check matrices are defined similarly with the elements of the matrices either defined over a ring $\mathbb{Z}_m$ or a finite field $\mathbb{F}_m$. 
The symplectic inner product with respect to the generalized Pauli basis is $\mathrm{symp}([\bm{a_i},\bm{b_i}],[\bm{a_j},\bm{b_j}]) = \bm{a_j}\bm{b_i}^T - \bm{a_i}\bm{b_j}^T$ and with respect to the finite field-based clock and shift operator is $\mathrm{symp}([\bm{a_i},\bm{b_i}],[\bm{a_j},\bm{b_j}]) = \mathrm{Tr}_{p^l/p}(\bm{a_j}\bm{b_i}^T - \bm{a_i}\bm{b_j}^T)$~\cite{Ketkar06,NBEASC2021}.

\subsection{CSS (Calderbank-Shor-Steane) codes}

Calderbank and Shor~\cite{CSS_CS}, and independently Steane~\cite{CSS_Steane}, proposed a framework to construct quantum error correction codes from two classical codes $C_1$ and $C_2$ that satisfy the dual-containing criterion $C_1^{\perp} \subset C_2$. 
The quantum codes constructed using this framework are called the \textit{CSS codes} and form a class of stabilizer codes. 
When the codes $C_1$ and $C_2$ used to construct a CSS code are the same, i.e., $C_1=C_2=C$, the code $C$ is a dual-containing classical code, i.e., $C^{\perp} \subset C$. 
Let $H_1$ and $H_2$ be the parity check matrices of the classical codes $C_1[n,k_1,d_1]$ and $C_2[n,k_2,d_2]$, respectively. 
When $C_1^{\perp} \subset C_2$, we obtain $H_2H_1^{\mathrm{T}} = \bm{0}$. 
The basis codewords of the CSS code $\mathcal{Q}_{\mathrm{CSS}}$ are the normalized superposition of all the elements in a particular coset of $C_1^{\perp}$ in $C_2$~\cite{NielsenChuang}. 
Thus, the CSS code obtained from $C_1$ and $C_2$ is an $\llbr n,(k_1+k_2-n), d \geq \mathrm{min}(d_1,d_2) \rrbr$ quantum code. 
The minimum distance is equal to $\mathrm{min}(d_1,d_2)$ when the code is non-degenerate~\cite{Nadkarni_TQE21_Qudit_CSS_Codes}, i.e., when the minimum Pauli weight of any stabilizer is at least the minimum distance of the code. 
The check matrix of the CSS code is 
\begin{align}
H_{\rm CSS} =
\left[\begin{array}{c|c}
    H_1 & 0 \\
    0 & H_2
\end{array}\right], 
\end{align}
where the $\mathrm{X}$- and $\mathrm{Z}$-stabilizers based on $C_1$ and $C_2$ correct the $\mathrm{Z}$- and $\mathrm{X}$-errors, respectively.
This is because the syndrome of an error $E(\bm{e}_{\mathrm{X}}, \bm{e}_{\mathrm{Z}})$ is 
\begin{align}
\bm{s} = 
\left[\begin{array}{c|c}
    H_1 & 0 \\
    0 & H_2
\end{array}\right]
\begin{bmatrix}
\bm{e}_{\mathrm{Z}}^T \\
\bm{e}_{\mathrm{X}}^T
\end{bmatrix}
=
\begin{bmatrix}
H_1 \bm{e}_{\mathrm{Z}}^T \\
H_2 \bm{e}_{\mathrm{X}}^T
\end{bmatrix} 
=
\begin{bmatrix}
\bm{s}_{\mathrm{Z}}^T \\
\bm{s}_{\mathrm{X}}^T
\end{bmatrix}.
\end{align}
Two independent decoders can be run for $H_1$ and $H_2$ with respective syndromes $\bm{s}_{\mathrm{Z}}$ and $\bm{s}_{\mathrm{X}}$ to correct the complete error.
While this is the commonly used strategy, when $\mathrm{X}$- and $\mathrm{Z}$-errors are correlated, it is suboptimal.

\subsection{Logical operators}

An $\llbr n,k,d \rrbr$ code encodes $k$ logical qubits into $n$ physical qubits such that the minimum (Pauli) weight of an undetectable error is $d$.
The undetectable errors are precisely the logical operators of the code since they non-trivially change the encoded information while keeping it within the code space.
This is similar to the classical case where undetectable errors are the codewords of the code --- adding a codeword to the transmitted codeword non-trivially changes the encoded message but keeps it within the code space.
Formally, we can identify the logical Pauli operators of the code with $\mathcal{N}(\mathcal{S})$, where the operators (stabilizers) within $\mathcal{S} \subseteq \mathcal{N}(\mathcal{S})$ are trivial logical operators as they leave any encoded state unchanged~\cite{Gottesman97 , wilde2009logical}.
Each logical qubit is identified by a pair of logical $\mathrm{X}$ and logical $\mathrm{Z}$ operators, denoted $\overline{\mathrm{X}}_j$ and $\overline{\mathrm{Z}}_j$ respectively for the $j^{\rm th}$ logical qubit, each of which act on the $n$ physical qubits of the code.
Naturally, $\overline{\mathrm{X}}_j, \overline{\mathrm{Z}}_j \in \mathcal{N}(\mathcal{S}) \setminus \mathcal{S}$ and satisfy the conditions
\begin{align}
\overline{\mathrm{X}}_i \, \overline{\mathrm{Z}}_j = 
\begin{cases}
- \overline{\mathrm{Z}}_j \overline{\mathrm{X}}_i & \text{if}\ i = j, \\
\overline{\mathrm{Z}}_j \overline{\mathrm{X}}_i & \text{if}\ i \neq j,
\end{cases}
\end{align}
for all $i,j \in \{ 1,2,\ldots,k \}$.
For a CSS code defined by classical codes $C_1$ and $C_2$ as explained before, the logical $\mathrm{X}$ (resp. logical $\mathrm{Z}$) operators are defined by the cosets in $C_2/C_1^{\perp}$ (resp. $C_1/C_2^{\perp}$) under the homomorphism $\gamma \colon \mathcal{P}_n \rightarrow \mathbb{F}_2^{2n}$~\cite{rengaswamy2020classical}.

As an example, consider the $\llbr 7,1,3 \rrbr$ \emph{Steane code}~\cite{CSS_Steane} defined by setting $C_1 = C_2 = C$ to be the classical $[7,4,3]$ Hamming code.
Then $H_1 = H_2 = H$ is the parity check matrix of the Hamming code.
The dual code $C^{\perp}$ is the $[7,3,4]$ binary simplex code which is a subcode of the Hamming code, i.e., $C^{\perp} \subset C$.
There is a single non-trivial coset in $C/C^{\perp}$ with coset leader $\bm{c} = 1111111$.
Hence, the single logical qubit has logical Pauli operator generators 
\begin{align}
\overline{\mathrm{X}} &= E(\bm{c},\bm{0}) = \mathrm{X}_1 \mathrm{X}_2 \mathrm{X}_3 \mathrm{X}_4 \mathrm{X}_5 \mathrm{X}_6 \mathrm{X}_7, \nonumber \\
\overline{\mathrm{Z}} &= E(\bm{0},\bm{c}) = \mathrm{Z}_1 \mathrm{Z}_2 \mathrm{Z}_3 \mathrm{Z}_4 \mathrm{Z}_5 \mathrm{Z}_6 \mathrm{Z}_7.    
\end{align}
It is easily verified that $\overline{\mathrm{X}} \, \overline{\mathrm{Z}} = - \overline{\mathrm{Z}} \, \overline{\mathrm{X}}$ since $\bm{c} \, \bm{c}^T = 1$ (mod $2$) guarantees that the symplectic inner product of $[\bm{c},\bm{0}]$ and $[\bm{0},\bm{c}]$ is $1$.
Since stabilizers do not modify the action of an operator, we can multiply $\overline{\mathrm{X}}$ or $\overline{\mathrm{Z}}$ with a weight-$4$ $\mathrm{X}$- or $\mathrm{Z}$-type stabilizer (from the rows of $H$), respectively, to reduce the weight to $3$.
Hence, the code has minimum distance $3$ because logical operators are non-trivial undetectable errors.

For universal computation on the logical qubits of a code, logical Pauli operators alone are insufficient.
It is necessary to synthesize a universal set of logical operators, such as $\overline{\mathrm{H}}_i, \overline{\mathrm{T}}_i$, and $\overline{\mathrm{CNOT}}_{i \rightarrow j}$ on all logical qubits $i,j \in \{ 1,2,\ldots,k \}$.
Besides correcting errors, this is an important aspect of QEC.
Typically, these logical operators are synthesized for individual codes or families of codes by leveraging their structural properties~\cite{Gottesman97 , Shor95 , horsman2012surface , litinski2019game , vuillot2019code , kubica2015universal , cohen2022low , krishna2021fault}.
There are systematic ways to approach this too for arbitrary codes~\cite{rengaswamy2020logical , rengaswamy2020classical}, but the resulting circuits might be suboptimal in terms of their circuit complexity.
It is essential to ensure that errors do not spread during the execution of these logical circuits.
This is the requirement of fault tolerance that we will discuss briefly later.
Construction of fault tolerant logical gates is unique to QEC and is critical for reliable and useful quantum computing.

\subsection{Quantum low-density parity check codes}

Quantum codes with high rate and good error correction capability are considered ideal candidates for fault-tolerant quantum computing (FTQC). 
In several architectures~\cite{XanaduPassiveArch}, the noise over the syndrome measurement scales with the number of qubits on which a stabilizer acts non-trivially and the number of stabilizers acting on a qubit. 
Thus, classes of quantum low-density parity check (QLDPC) codes~\cite{breuckmann2021quantum} with asymptotically constant rate and distance scaling linearly with $n$ are preferred candidates for FTQC.  

Surface codes~\cite{dennis2002topological , fowler2012surface} are well-studied $\llbr n, \mathcal{O}(1), \mathcal{O}(\sqrt{n}) \rrbr$ QLDPC codes that have good logical error rate performance, good distance scaling, and require only nearest-neighbor connectivity of qubits. 
However, they have asymptotically zero rate, leading to large overheads as the size of the system increases. 
There have been efforts in moving beyond the surface code and exploring codes with constant non-zero asymptotic rates and with distance scaling linearly with code size. 
The \emph{hypergraph product} codes~\cite{tillich2013quantum} and the \emph{lifted product} codes~\cite{Panteleev2021} are two important classes of CSS codes that are currently being explored, besides other codes~\cite{breuckmann2021quantum , leverrier2022quantum , mostad2024generalizing}. 
We provide a brief review of surface codes, hypergraph product codes, lifted product codes, and concatenated quantum codes (which are more general than QLDPC codes).

\subsubsection{Surface codes and Toric codes}
 
\begin{figure}
    \centering
    \includegraphics[scale=0.3,keepaspectratio]{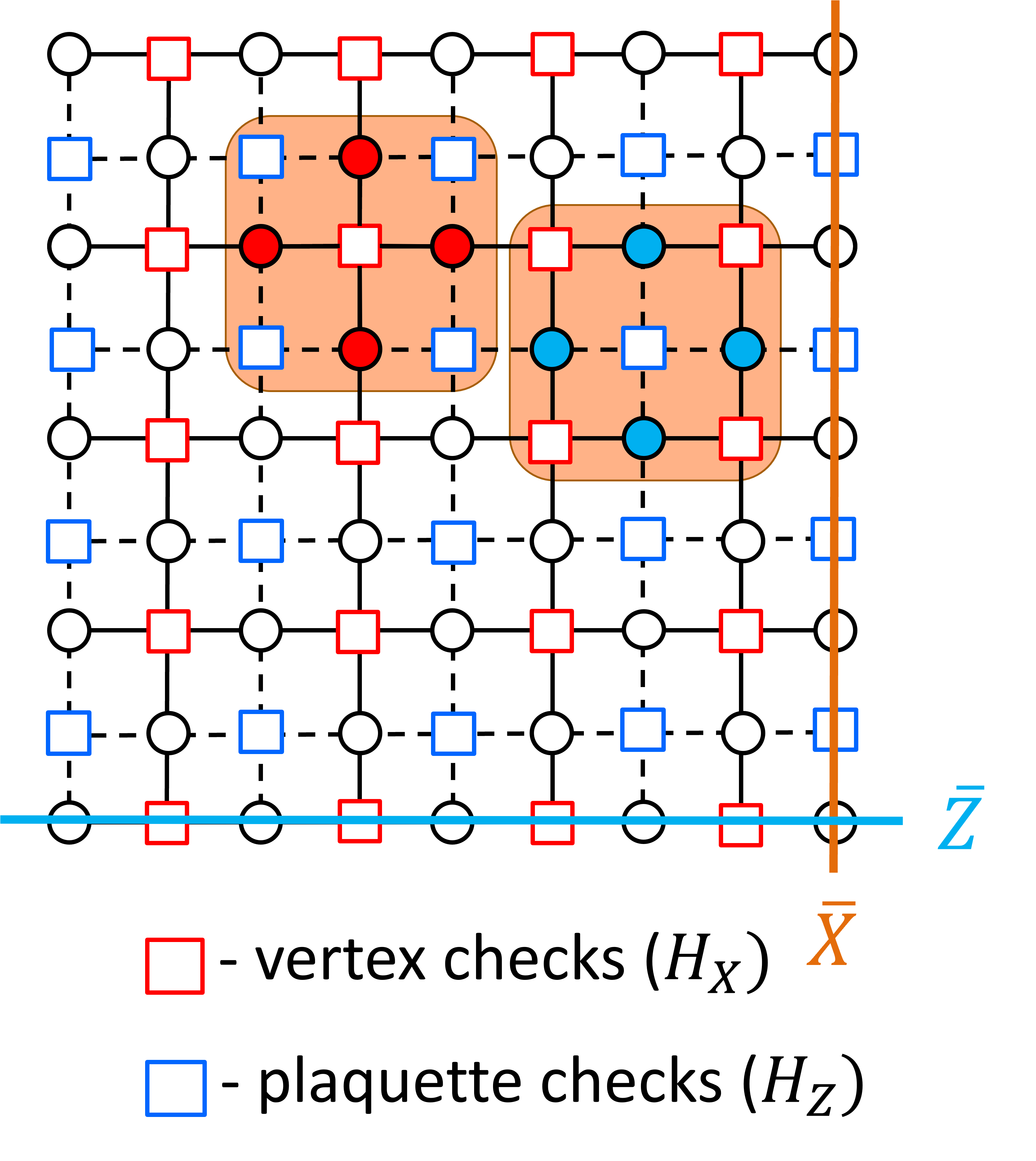}
    \caption{The standard $\llbr 41,1,5 \rrbr$ surface code with side length $5$. Qubits are represented by circles. The orange boxes highlight weight-$4$ $\mathrm{X}$-type and $\mathrm{Z}$-type stabilizers represented by red and blue squares, respectively. Note that there are weight-$3$ stabilizers in the boundaries of the lattice. All connecting lines represent local connectivity natively made available in the hardware; solid lines show the faces (or plaquettes) of the lattice and dashed lines connect the blue $\mathrm{Z}$-checks to qubits incident to a face. A possible choice for the logical $\mathrm{Z}$ and logical $\mathrm{X}$ operators is shown, both of minimum weight $5$.}
    \label{fig:surface_code_5}
\end{figure}

The surface codes are well-studied QLDPC codes whose qubits can be laid out on a 2D surface and the stabilizer measurements involve qubits within a particular neighborhood on the surface. 
Thus, they require only nearest-neighbor connectivity of qubits, which is essential for superconducting architectures. 
The most studied surface code is defined over a square lattice with qubits represented by edges, $\mathrm{X}$-stabilizers represented by vertices, and $\mathrm{Z}$-stabilizers represented by faces. 
A $\mathrm{Z}$-stabilizer acts non-trivally on all the qubits defined by the edges defining the face. 
An $\mathrm{X}$-stabilizer acts non-trivially on all the qubits incident to the vertex representing the stabilizer. 
The logical $\mathrm{X}$ and $\mathrm{Z}$ operators span the length and breadth of the lattice and the distance of the code is based on the side length of the square lattice. 
This standard surface code is shown in Fig.~\ref{fig:surface_code_5}.

The toric code~\cite{kitaev2003fault} is defined on a square lattice on the surface of a torus with the edges of the lattice depicting qubits, the faces representing $\mathrm{Z}$-stabilizers, and vertices representing $\mathrm{X}$-stabilizers. 
The logical operators are represented by topologically non-trivial loops in the lattice, which also span the length and the breadth of the lattice. 
The toric code encodes two logical qubits and has distance based on the side length of the lattice. 
The square lattice for the toric code is the same as for the surface code but with opposite boundaries identified with each other, i.e., it has no boundaries.

\subsubsection{Hypergraph product codes}

\begin{figure}
    \centering
    \includegraphics[scale=0.3,keepaspectratio]{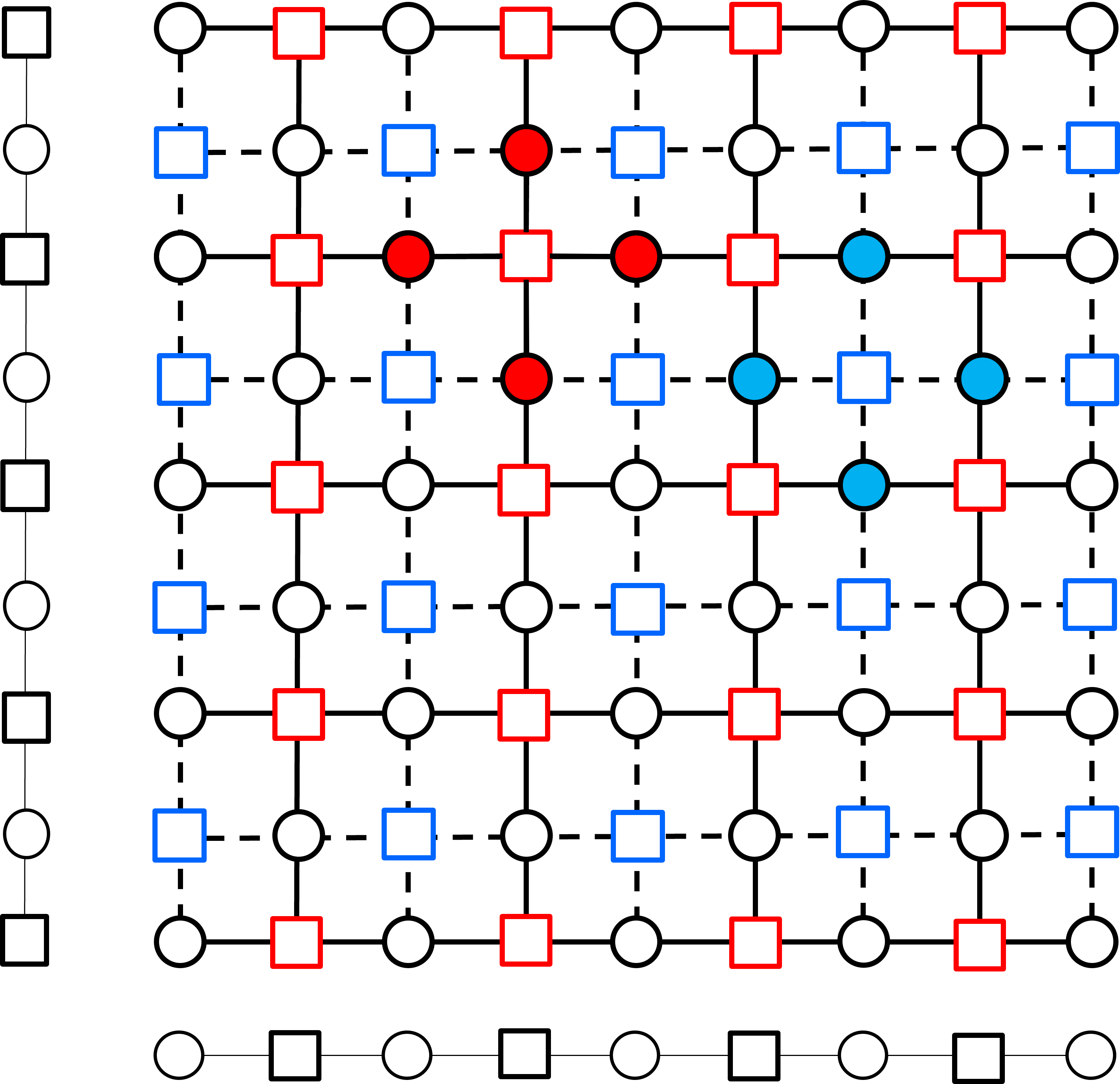}
    \caption{The surface code constructed as the hypergraph product of classical repetition codes. The intersection of bits and checks of the repetition codes determine the qubits and stabilizers of the hypergraph product code.}
    \label{fig:surface_code_hgp}
\end{figure}
 
Tillich and Zémor \cite{tillich2013quantum} proposed the hypergraph product (HGP) code contruction based on any two classical codes, called the component codes. 
The Tanner graph of the HGP code is the graph product of the Tanner graphs of the component codes. 
For $i=1,2$, let $C_i[n_i,k_i,d_i]$ be the two component codes with parity check matrix $H_i$. 
Let their tranposed codes $C_i^T$ be the code with parity-check matrix $H_i^T$ with parameters $[m_i, k_i^T, d_i^T]$. 
The hypergraph product code $HGP(C_1,C_2)$ obtained is a $[[n_1n_2+m_1m_2, k_1k_2 + k_1^Tk_2^T, \mathrm{min}(d_1,d_2,d_1^T,d_2^T)]]$ CSS code with $\mathrm{X}$- and $\mathrm{Z}$-check matrices
\begin{align}
H_{\mathrm{X}} &= \begin{bmatrix}H_1 \otimes I & I \otimes H_2^T\end{bmatrix}, \nonumber \\ H_{\mathrm{Z}} &= \begin{bmatrix} I \otimes H_2 & H_1^T \otimes I\end{bmatrix}. 
\end{align}
When the component codes are chosen appropriately, the parameters follow the scaling $\llbr n, \mathcal{O}(n), \mathcal{O}(\sqrt{n}) \rrbr$.
The surface code can be constructed as the hypergraph product of classical repetition codes as shown in Fig.~\ref{fig:surface_code_hgp}.
There are closed-form expressions for the logical Pauli operators of these codes~\cite{burton2021limitations}.

\subsubsection{Lifted product codes}

Panteleev and Kalachev \cite{Panteleev2021 , Panteleev2022 , panteleev2022asymptotically} first proposed the lifted product (LP) codes, which are the lifted versions of the hypergraph product codes. 
The LP codes are based on two matrices $A_1$ and $A_2$ defined over $R_l = \mathbb{F}_2[x]/(x^l-1)$. 
Let $A_i$ be of size $(m_i\times n_i)$. 
For $a(x) = a_0 + a_1x + \cdots + a_{l-1}x^{l-1} \in R_l$, the lift $\mathbb{B}(a(x))$ is the cyclic permutation matrix whose first column has the coefficients of $a(x)$ and the rest of the columns are obtained as the previous columns shifted down by one element. 
We note that  $a^T(x) = a_0 + a_{l-1}x + \cdots + a_{1}x^{l-1}$ . 
The lifted product code $LP(A_1, A_2)$ is the $l(n_1m_2+n_2m_1)$-qubit CSS code with $\mathrm{X}$- and $\mathrm{Z}$-check matrices
\begin{align}
H_{\mathrm{X}} &= \mathbb{B}([A_1 \otimes I~~I \otimes A_2]), \nonumber \\ 
H_{\mathrm{Z}} &= \mathbb{B}([I \otimes A_2^T ~~ A_1^T \otimes I]). 
\end{align}
When the component codes are chosen appropriately, the parameters follow the scaling $\llbr n, \mathcal{O}(n), \mathcal{O}(n) \rrbr$~\cite{panteleev2022asymptotically}.

\subsubsection{Concatenated quantum codes}

Concatenated quantum codes are obtained by concatenating a quantum code, called the inner code, with another quantum code, known as the outer code~\cite{knill1996concatenated , jochym2014using , yoshida2024concatenate}. 
The inner code is first used to encode a $K_1$-dimensional quantum system into $N_1$-dimensional quantum states. 
The $N_1$-dimensional quantum states are further encoded using an outer code by considering the $N_1$-dimensional quantum states as logical information. 
The distance of the concatenated code is the product of the distance of the outer and inner code, improving its error correction ability. 
Multiple outer codes could be used to encode the logical information over the $N_1$-dimensional quantum states.
The quintessential example of a concatenated code is the $\llbr 9,1,3 \rrbr$ \emph{Shor code}, which was constructed famously by Peter Shor~\cite{Shor95} to show that QEC even works.
Until then, the continuous nature of quantum errors was thought to be a fundamental bottleneck to construct reliable quantum systems.
The Shor code first encodes a single qubit in the $\llbr 3,1,1 \rrbr$ phase flip code defined by the stabilizer group $\mathcal{S}_{\rm phase} = \langle \mathrm{X}_1 \mathrm{X}_2, \mathrm{X}_2 \mathrm{X}_3 \rangle$ and logical operators $\overline{\mathrm{X}} = \mathrm{X}_1, \overline{\mathrm{Z}} = \mathrm{Z}_1 \mathrm{Z}_2 \mathrm{Z}_3$.
Then it encodes each of those $3$ qubits into the bit flip code defined by the stabilizer group $\mathcal{S}_{\rm bit} = \langle \mathrm{Z}_1 \mathrm{Z}_2, \mathrm{Z}_2 \mathrm{Z}_3 \rangle$ and logical operators $\overline{\mathrm{Z}} = \mathrm{Z}_1, \overline{\mathrm{X}} = \mathrm{X}_1 \mathrm{X}_2 \mathrm{X}_3$.
Overall, the concatenated code has the stabilizer group
\begin{align}
\mathcal{S}_{\rm Shor} &= \langle \ \mathrm{Z}_1 \mathrm{Z}_2 \ , \mathrm{Z}_2 \mathrm{Z}_3 \ , \mathrm{Z}_4 \mathrm{Z}_5 \ , \ \mathrm{Z}_5 \mathrm{Z}_6 \ , \mathrm{Z}_7 \mathrm{Z}_8 \ , \ \mathrm{Z}_8 \mathrm{Z}_9 \ , \nonumber \\
  & \qquad \mathrm{X}_1 \mathrm{X}_2 \mathrm{X}_3 \mathrm{X}_4 \mathrm{X}_5 \mathrm{X}_6 \ , \ \mathrm{X}_4 \mathrm{X}_5 \mathrm{X}_6 \mathrm{X}_7 \mathrm{X}_8 \mathrm{X}_9 \ \rangle.
\end{align}
A valid pair of logical Pauli operators are 
\begin{align}
\overline{\mathrm{X}} = \mathrm{X}_1 \mathrm{X}_2 \cdots \mathrm{X}_9 \ , \  \overline{\mathrm{Z}} = \mathrm{Z}_1 \mathrm{Z}_2 \cdots \mathrm{Z}_9.    
\end{align}
Clearly, by multiplying with stabilizers, one can make them weight-$3$ so that the minimum distance of the code is $3$.

\subsection{Bosonic codes}

Bosonic encoding, also known as continuous-variable encoding, encodes quantum information into electromagnetic signals. 
Bosonic encoding can be viewed analogous to the modulation codes used in communication systems where bitstrings are encoded into the in-phase and quadrature carrier electromagnetic waves.

Bosonic codes are classified as bosonic stabilizer codes and bosonic Fock-state codes. The bosonic encodings inherently have error correction ability embedded into them. 
The logical performance of qubit codes can be improved by concatenating them with bosonic codes~\cite{noh2020fault , noh2022low , grimsmo2021quantum , raveendran2022finite}. 
The hardware can be utilized more efficiently using these codes and certain gates forbidden over qubits can be performed using the continuous-variable operations.

In bosonic stabilizer encoding, the carrier electromagnetic waves correspond to the position and momentum quadratures.
Gottesman-Kitaev-Preskill (GKP) encoding~\cite{gottesman2001encoding} is the most commonly used bosonic stabilizer encoding. 
A commuting set of displacement operators across position and momentum quadratures form the stabilizers of the code.
A square GKP encoding can be viewed as a comb of evenly spaced momentum states with a spacing of $2\sqrt{\pi}$. 
Thus, a displacement of $2\sqrt{\pi}$ corresponds to a stabilizer and a displacement of $\sqrt{\pi}$ is a logical operation that transforms logical $\ket{0}$ to logical $\ket{1}$. The GKP code protects the quantum information from large displacements upto $\sqrt{\pi}/2$. 
Using the concepts of QEC with Fock states or number states, cat encoding, binomial encoding, rotor GKP encoding etc. are developed. 
The square GKP encoding and rotor encoding can be viewed to be analogous to amplitude-shift and phase-shift keying techniques, respectively. 
See~\cite{BosonicEncodingReview} for an extensive review of bosonic encoding.

\subsection{Decoding of quantum codes}

The decoder for quantum codes is still a classical algorithm that takes the quantum check matrix $H_{\mathcal{S}}$ and the syndrome as input and outputs an estimate of the error that caused the syndrome.
While the principle is similar to decoding of classical codes, there are multiple equivalent errors for the same syndrome due to degenerate stabilizer errors.
The optimal decoder on the depolarizing channel is not the maximum likelihood decoder but the maximum likelihood \emph{coset} decoder that determines the most likely logical coset $E + L + \mathcal{S}$ in $\mathcal{P}_n$ that matches the syndrome~\cite{pelchat2013degenerate , iyer2015hardness , fuentes2021degeneracy}.
Here, $E$ is the actual error and $L$ is a logical Pauli operator.
For fixed $E$ and $L$, all elements of the coset $E + L + \mathcal{S}$ have the same effect on the code since elements of $L + \mathcal{S} \in \mathcal{N}(\mathcal{S})$ have a trivial syndrome.
As discussed earlier, if $\hat{E}$ is the error estimate from a decoder (that has the same syndrome as $E$), then there are two possible scenarios: $\hat{E} E \in \mathcal{S}$ (correct decoding) or $\hat{E} E = L \in \mathcal{N}(\mathcal{S}) \setminus \mathcal{S}$ (logical error).
The plot of logical error rate versus noise parameter shows the block error rate performance of the code-decoder pair, just as in classical channel coding.

For CSS codes, one can execute separate decoders for $\mathrm{X}$-errors and $\mathrm{Z}$-errors using $H_{\mathrm{Z}}$ and $H_{\mathrm{X}}$, respectively.
In particular, QLDPC codes can be decoded using efficient message passing algorithms such as belief propagation or min-sum executed on the Tanner graphs of $H_{\mathrm{Z}}$ and $H_{\mathrm{X}}$~\cite{poulin2008iterative , babar2015fifteen , roffe2020decoding , Panteleev2021 , du2022stabilizer , du2023layered}.
Message passing can also be performed in the GF(4) domain by constructing a combined Tanner graph that includes all the stabilizers~\cite{kuo2020refined}.
In either case, short cycles and trapping sets, especially uniquely quantum ones from degeneracy~\cite{raveendran2021trapping , pradhan2023learning}, cause challenges in effective decoding that remain to be addressed.
Many families of QLDPC codes have a \emph{threshold}, which is the noise parameter beyond which increasing the code size within the family monotonically improves logical error rate on one side of the threshold and worsens logical error rate on the other side.
While QEC theorists strive to improve the threshold, experimentalists work hard to reduce the noise parameter as much as possible.
The \emph{threshold theorem}~\cite{aharonov1997fault , NielsenChuang} states that if every component in the hardware has an error rate within the threshold, then scalable and reliable quantum computers can be built through appropriate QEC schemes.

\subsection{Fault tolerance}

While decoding ensures that the most likely errors are corrected, quantum computers must also perform computation on the encoded information.
The circuits used to perform such universal computation must not spread errors catastrophically and overwhelm the QEC scheme.
This is ensured by imposing fault tolerance constraints on the circuits~\cite{shor1996fault , gottesman1998theory , steane1999efficient}.
There are several ways to define the requirement of fault tolerance, so let us consider a common one~\cite{gottesman2010introduction , gottesman2013fault , chamberland2018flag}.
Assume that every logical operator circuit is followed by a block of ideal syndrome measurement and error correction.
If the code can correct Pauli errors on up to $t$ qubits, then fault tolerance can be ensured by requiring that any combination of $t$ faults in the input and the logical operator circuit does not cause more than $t$ errors at the output of the circuit.
Of course, the syndrome measurement and error correction block could itself introduce noise.
But this is captured as errors in the input of the next logical operator circuit.
Since iterative decoders do not necessarily correct up to $t$ errors, fault tolerance can take a more subtle role.
Constructing fault tolerant gates on good QLDPC codes is an active and open area of research today.

Typically, logical Clifford gates are easier to construct on quantum codes than logical non-Clifford gates.
There are a variety of methods to construct logical Clifford gates.
It is common to design codes where a \emph{transversal} physical operation induces the necessary action on the logical qubits.
A transversal gate is one that acts as a tensor product on the $n$ physical qubits, thereby not introducing any interaction between two or more qubits~\cite{NielsenChuang}.
Hence, by design, it is a fault-tolerant circuit.
For example, the Steane code realizes the logical $\mathrm{H}$ and $\mathrm{S}$ transversally as $\mathrm{H}^{\otimes 7}$ and $\mathrm{S}^{\otimes 7}$, respectively.
If two logical qubits are encoded separately in two Steane code blocks, then a transversal $\mathrm{CNOT}$, i.e., $7$ $\mathrm{CNOT}$s between corresponding qubits of the two blocks, realizes the logical $\mathrm{CNOT}$.
However, the logical $\mathrm{T}$ gate cannot be realized transversally on the Steane code to complete a universal logical gate set.
In fact, the \emph{Eastin-Knill theorem}~\cite{eastin2009restrictions} states that no error-detecting quantum code can realize a universal logical gate set using only transversal gates.

An innovative strategy to implement logical non-Clifford gates is using \emph{magic states}~\cite{bravyi2005universal}.
These are specific resource states that enable one to implement the gate without directly applying it on the data qubits.
For example, the magic state for the $\mathrm{T}$ gate is $\ket{\mathrm{T}} = \frac{\ket{0} + e^{\mathrm{i} \pi/4} \ket{1}}{\sqrt{2}}$.
Given this state, the following circuit applies the $\mathrm{T}$ gate on the input data qubit $\ket{\psi}$ using \emph{only Clifford operations and Pauli measurements}:
\begin{center}
\begin{quantikz}
\lstick{$\ket{\mathrm{T}} = \mathrm{T} \ket{+}$} & \ctrl{1} &  & \gate{\mathrm{SX}} & \rstick{$\mathrm{T} \ket{\psi}$} \\
\lstick{$\ket{\psi} = \alpha \ket{0} + \beta \ket{1}$} & \targ{} & \meter{\mathrm{Z}} & \ctrl[vertical wire=c]{-1}\setwiretype{c}
\end{quantikz}
\end{center}
The double wire indicates a classically-controlled $\mathrm{SX}$ gate which is applied if and only if the measurement result is $-1$.
Hence, it is desirable to generate $\mathrm{T}$ magic states of high fidelity.
This is achieved through a process called \emph{magic state distillation (MSD)}~\cite{bravyi2005universal}.
The most common approach to MSD is the Bravyi-Haah protocol using \emph{triorthogonal codes}~\cite{bravyi2012magic}.
These codes realize logical $\mathrm{T}$ gates on all $k$ logical qubits via a transversal $\mathrm{T}$ gate on the $n$ physical qubits.
Once these codes are used to distill higher-fidelity magic states from lower-fidelity magic states natively produced by hardware, the resulting states are injected into the data using the above circuit.
When the data is itself encoded into a different code, such as a QLDPC code, the magic state must also be encoded to allow a fault-tolerant execution of the above circuit.
A major portion of the resource consumption of a quantum computer comes from magic state distillation and injection, since $\mathrm{T}$ gates are a critical component of most non-trivial quantum algorithms~\cite{litinski2019magic , chamberland2020very}.
Triorthogonal codes have been generalized to \emph{CSS-T} codes~\cite{rengaswamy2020optimality , rengaswamy2020csst} in the hope of reducing the overhead of implementing logical non-Clifford gates.
This has generated much interest among algebraic coding theorists recently~\cite{andrade2023css , berardini2024structure , camps2024algebraic , camps2024toward}.
The realization of logical non-Clifford gates with low overhead on good QLDPC codes is an important and exciting area of research.

\section{Conclusion}

In this short article, we have briefly reviewed the fundamentals of quantum computation and quantum error correction.
We hope that this is informative to researchers that are new to the field.
There are several challenges to be addressed in the pursuit of scalable, fault-tolerant, quantum computing.
We firmly believe that classical coding theorists have a lot to offer in addressing these challenges.

\IEEEtriggeratref{94}



\end{document}